\documentclass[%
 aip,
 jmp,%
 amsmath,amssymb,
 reprint,%
]{revtex4-1}

\usepackage{graphicx}% Include figure files
\usepackage{dcolumn}% Align table columns on decimal point
\usepackage{bm}% bold math
\usepackage[mathlines]{lineno}% Enable numbering of text and display math
\usepackage{textcomp}
\begin{document}

\title{A compact broadband terahertz range quarter-wave plate}%

\author{Liang Wu} 
\email{liangwu.jhu@gmail.com}
\altaffiliation{New address: Department of Physics and Astronomy, University of Pennsylvania, Philadelphia, Pennsylvania 19104, USA}
\affiliation{Department of Physics and Astronomy, The Johns Hopkins University, Baltimore, Maryland, 21218, USA}
\author{A. Farid} 
\affiliation{Department of Physics and Astronomy, The Johns Hopkins University, Baltimore, Maryland, 21218, USA}
\author{N. J. Laurita} 
\affiliation{Department of Physics and Astronomy, The Johns Hopkins University, Baltimore, Maryland, 21218, USA}
\author{T. Mueller} 
\affiliation{Department of Materials Science and Engineering, The Johns Hopkins University, Baltimore, Maryland, 21218, USA}
\author{N. P. Armitage}
\affiliation{Department of Physics and Astronomy, The Johns Hopkins University, Baltimore, Maryland, 21218, USA}
\date{\today}

\begin{abstract}

We detail the design and characterization of a terahertz range achromatic quarter-wave plate based on a stack of aligned variable thickness birefringent sapphire discs.   The disc thicknesses and relative rotations of the discs are determined through a basin-hopping Monte Carlo thermal annealing routine.  The basin-hopping scheme allows an improved refinement of the required thicknesses and rotations to give a predicted phase error from the ideal $\pi/2$ of only $0.5 \%$, which is a factor of approximately 6 better than previous efforts.  Moreover, the large contrast between extraordinary and ordinary axes of sapphire allow us to greatly decrease the overall optical path length of our wave plate design by approximately a factor of 10 over similar designs based on quartz discs.  However, this very same contrast requires very precise tolerances in the milled thicknesses of the discs and their assembly.   We detail a method to compensate for differences in the thickness from their calculated ideal values.   We have constructed one of our designs and found it similar in performance to other configurations, but with our much more compact geometry.

\end{abstract}
                             

\maketitle

\section{\label{Introduction}Introduction}

THz frequencies and picosecond time scales are found ubiquitously in condensed matter physics and material systems.  In recent years,  the use of the technique of time-domain terahertz spectroscopy (TDTS) \cite{nuss1998terahertz} has grown dramatically and been employed in many different areas including materials science \cite{kaindl2003ultrafast,heyman1998time,richter2010exciton}, monitoring of pharmaceutical ingredients \cite{taday2004applications}, security and biohazard detection \cite{choi2004potential,leahy2007wideband}, structural and medical imaging\cite{hu1995imaging,johnson2001enhanced,mittleman1997t,wang2003t,chan2007imaging} and the study of materials with low temperature quantum correlations \cite{corson1999vanishing,bilbro2011temporal,aguilar2012terahertz,aguilar2013aging,wu2013sudden,hancock2011surface,pan2014low,morris2014hierarchy,laurita2015singlet,bosse2014a,little2017antiferromagnetic}.  In the TDTS technique, an approximately one $ps$ long pulse of electromagnetic radiation is transmitted through the sample.    Use of such a time limited pulse gives multiplexing advantages, as it is composed of a large range of frequency components and therefore gives the ability to do broadband spectroscopy in a single source and detection scheme.  Although this has obvious advantages, it can complicate measurements because many THz and millimeter range optical components are narrow band.

In the characterization of materials, THz range polarimetry is becoming increasingly important.   Many materials show strong polarization anisotropies that provide important information about the magnetic and electronic states of the materials under test \cite{Xia06a,Qi2008,Nandkishore2012,Tse11a,Maciejko10a,Armitage14a}.   Recent advances allow for precise polarimetry and make measurements possible that could not be made previously \cite{Castro-Camus2005,Makabe2007,Shimano2011,Neshat12a,aguilar2012terahertz,MorrisOE12,George12a,koirala2015record, wu2015high, wu2016tuning, wu2016quantized, laurita2017anomalous}.  Despite these advances in \textit{measuring} polarization states it is still surprisingly challenging to reliably \textit{generate} TDTS radiation of a non-trivial polarization state.   Part of the issue here is the general difficulty in manipulating THz range radiation.  However, an additional difficulty arises due to the broadband nature of a TDTS pulse and the intrinsically narrow band nature of typical devices like conventional multi-order $\frac{1}{4}$-wave plates.

Achromatic wave plates in the visible range have a long history.   In pioneering work, West and Makas \cite{West49a} combined plates with different birefringent dispersions to achieve partial cancelation of the phase shift giving a large portion of the visible frequency range with the same $\pi/2$ phase shift.   This is how essentially all commercial achromatic wave plates are now made.  Destriau and Prouteau \cite{Destriau49a} combined a half-wave plate and a quarter-wave plate of the same material and succeeded in creating a new effective quarter-wave plate that extended the bandwidth of the composite retardation plate into the entire visible part of the electromagnetic spectrum.   Much more recently, Masson and Gallot \cite{Masson06a} succeeded in extending these ideas to a multi-layers structure with thicknesses appropriate to the THz regime.   They optimized the thicknesses and relative orientation of 6 birefringent quartz plates through a thermal annealing algorithm to achieve a phase delay of $\pi/2 \pm 3\%$ over a frequency range of 0.25 - 1.75 THz.  Although, the device of Ref. \onlinecite{Masson06a} is remarkable with its broad bandwidth, because it uses quartz which has an overall weak frequency birefringent contrast ($n_o = 2.108$, $n_e =  2.156$) between fast and slow axes, the structure has an overall length of almost 3 cm.  A long optical path length as such makes its incorporation in usual terahertz setups difficult without realignment.

In the present work, we have succeeded in designing and implementing an multi-plate arrangement based on sapphire discs.   The much larger contrast between optical axes in sapphire as compared to quartz means that the device reduces the optical path length by an order of magnitude over previous designs (approximately  3 mm in our design), which makes it far easier to incorporate into an existing optical setup to achieve circularly polarized light.   The larger birefringent contrast and compact size of the insertion device comes at the cost of much more precise required tolerances with respect to angular misalignment and machining thicknesses.  In this work we use a basin-hopping Monte Carlo algorithm to determine the optimal thicknesses and relative rotations.   A few different potential scenarios are discussed, but in all cases the predicted phase error can be kept below $0.5\%$.     Because of the small overall thicknesses, the device with sapphire discs is very sensitive to misalignment and machining tolerances.  We detail a procedure that minimizes the problems associated with such errors.    We have assembled one of these devices and for one of the scenarios we found an overall phase error of a few percent in the 0.1  - 0.8 THz range.

\section{\label{Methods}Methods}

As mentioned above, in contrast to other recent efforts to implement multi-layers structure for THz waveplates, we use sapphire due to large difference in index of refraction between its ordinary and extraordinary axes.   This allows the device to be made very compact and easily incorporated into existing spectroscopic systems.   Sapphire (Al$_2$O$_3$) was chosen because it is a conventional optical window material with a large birefringent contrast ($n_o = 3.39$, $n_e =  3.07$), which also has very low absorption losses in the THz range.

As done in Ref. \onlinecite{Masson06a} we combined a stack of birefringent plates with specified thicknesses and rotations such as to achieve the smallest overall deviation from a $\pi/2$ phase shift in the target spectral range.   The optical properties of an individual layer $i$ is computed from its Jones matrix which for a wave plate with phase delay $\delta_i$ and relative rotation angle $\theta_i$.

\begin{widetext}
\begin{equation}
J_i(\delta_i,\theta_i) = 
\left[\begin{array}{cc} \mathrm{cos \,  \,} \delta_i/2 + i \mathrm{cos \, } 2\theta_i  \; \mathrm{sin \, } \delta_i/2 & i \mathrm{sin \, } 2\theta_i  \;\mathrm{sin \, } \delta_i/2 \\  i \mathrm{sin \, } 2\theta_i  \; \mathrm{sin \, } \delta_i/2 & \mathrm{cos \, } \delta_i/2 - i \mathrm{cos \, } 2\theta_i   \; \mathrm{sin \, } \delta_i/2  \end{array}\right] .
\label{ABCD}
\end{equation}
\end{widetext}

Due to the symmetry properties of waveplates \cite{Armitage14a} the total Jones matrix for the full stack is given by

\begin{equation}
J = \prod _i J_i = 
\left[\begin{array}{cc}A & B \\- B^* & A^* \end{array}\right] .
\end{equation}

The resulting phase shift between principle axes of the wave plate is

\begin{equation}
 \delta =2\mathrm{arctan} \,  \sqrt {\frac{|\mathrm{Im} A|^2 +|\mathrm{Im} B|^2  }{|\mathrm{Re} A|^2 +|\mathrm{Re} B|^2}}
%\mathrm{tan}^2 \, \frac{\delta}{2} =  \frac{|\mathrm{Im} A|^2 +|\mathrm{Im} B|^2  }{|\mathrm{Re} A|^2 +|\mathrm{Re} B|^2}
 \end{equation}

We performed a basin-hopping Monte Carlo simulations \cite{wales1997global} to optimize the layer thickness and rotation by minimizing the error function $\Sigma_\omega[\delta(\omega) - \pi/2]^2$ over the frequency range 0.1 - 2 THz.   Basin-hopping is an algorithm for efficient global optimization for finding minima.  It has the advantage over conventional thermal annealing because it allows one to sample a larger parameter space so as not to get stuck in local minima of the error function.

\section{\label{ResultsDiscussion}Results and Discussion}

\begin{figure}[tbp]
\centering
\includegraphics[width=8.5cm]{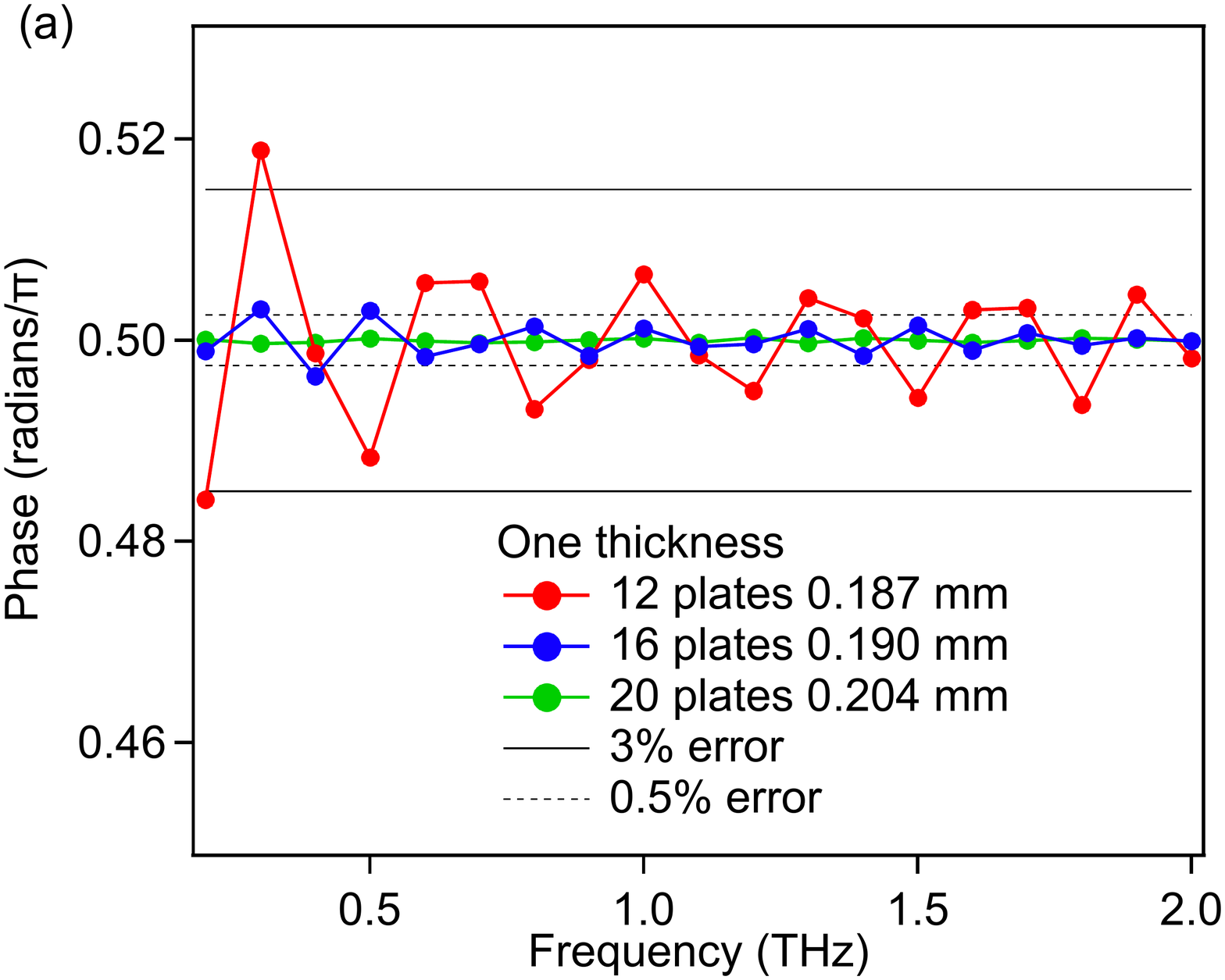}
\includegraphics[width=8.5cm]{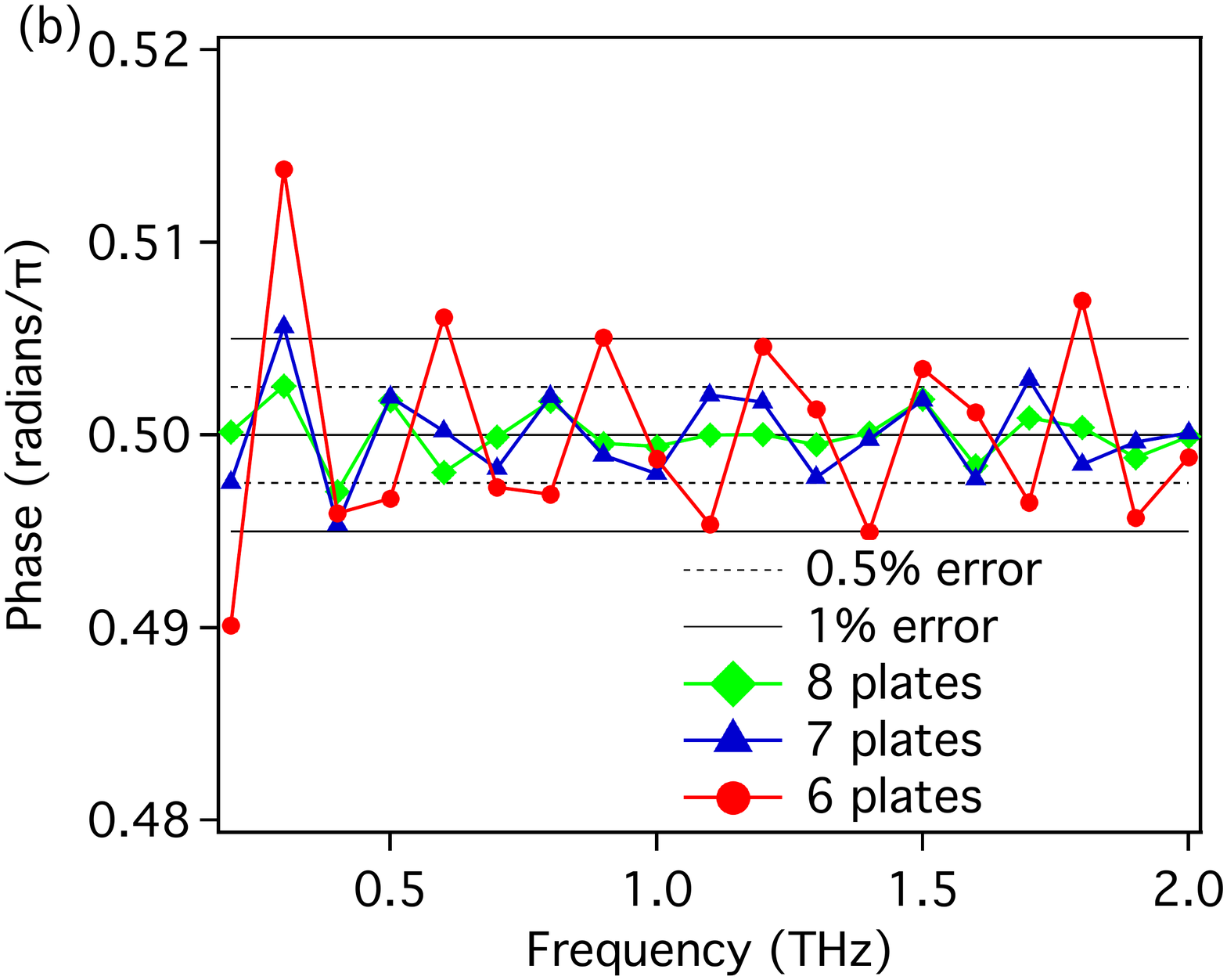}
\caption{(a) Simulated phase difference between $x$ and $y$ directions for the wave plate configurations with one thickness (b) Simulated phase difference between $x$ and $y$ directions for the wave plate configurations with 6,7, and 8 plates using two possible thicknesses (configurations A, B, and C) given in Table I.  }
\label{Simulation1}
\end{figure}

We explored a number of different possible configurations with different combinations of sapphire plates.   In all cases the overall length of the device was approximately 2-4 mm, which is significantly reduced from the device of Ref. \onlinecite{Masson06a} which was approximately 3 cm long.  Although in principle one can also choose to optimize the thickness for every layer also, here we chose to allow the algorithm to optimize with one or two thicknesses as ultimately having only a few possible different kinds of custom optical components manufactured greatly reduces machining costs while still achieving the same accuracy in phase.  Ultimately we believe we are dominated by alignment errors in assembly and machining thickness so choosing a discrete number of possible thickness does not introduce significant additional errors and is more cost effective. In Fig. \ref{Simulation1} (a) we show configurations for different numbers of plates with a single thickness that gives errors 1-3 $\%$.  One can see in Fig. \ref{Simulation1} (b) for three different configurations (A, B, C) of a different number of plates with two unique thicknesses.   The typical error in all cases was approximately $0.5\%$, which was significantly reduced from combinations with a single thickness.   It is also reduced from combinations arrived at without basin-hopping algorithms which were typically about $3\%$.   In principle, all these combinations are suitable for assembly depending on one's resources and needs.     The parameter values for three different configurations shown in Fig. \ref{Simulation1} (b) can be found in Table I.

\begin{table*}
\begin{ruledtabular}
\begin{tabular}{ccccccccccccccccc}
& &1 &2 &3
 & 4 & 5 & 6 & 7 & 8 & Total &  $ \Sigma_\omega[\frac{\delta(\omega)}{\pi/2} - 1]^2$ \\
\hline
A& Thickness (mm) &0.5660&0.3774&0.3774&0.3774&0.3774&0.3774& &  & 2.453 & 5.5 $\times$ 10$^{-3}$\\
& Angle (degrees) & 12.3 & 56.7 & 28.1 
& 356.5 & 302.0 & 349.2 & &  \\
\hline
B & Thickness (mm) & 0.5687 & 0.3791 & 0.3791 &0.3791 &0.3791 &0.3791 & 0.5687 &  & 3.032 & 1.0 $\times$ 10$^{-3}$\\
 & Angle (degrees) & 21.9 & 66.4 & 4.3
& 323.4 & 337.7 & 299.2 &357.3 &  \\
\hline
C &Thickness (mm) &0.5701 &0.3798 & 0.3798 &0.3798 &0.3798 &0.3798&0.3798 & 0.5701 & 3.419 & 3.4 $\times$ 10$^{-4}$\\
 & Angle (degrees) & 303.9 &263.4& 330.2 
& 344.9 & 29.8 & 15.4 &308.3 & 347.7  \\

\end{tabular}
\end{ruledtabular}

\caption{\label{tab:table2} Calculated Thickness and Orientations of Sapphire Plates for THz $1/4$ wave plate.   Numbers across the top represent each plate in the stack.  Total represents the total thickness in mm.   Phase error is given as the total accumulated phase error as a sum over frequency points with a step of 0.05 THz from 0.2 THz to 2 THz, normalized by $\pi/2$.  }

\end{table*}

\begin{figure}[tbp]
\centering

\includegraphics[angle = -0,width=8.5cm]{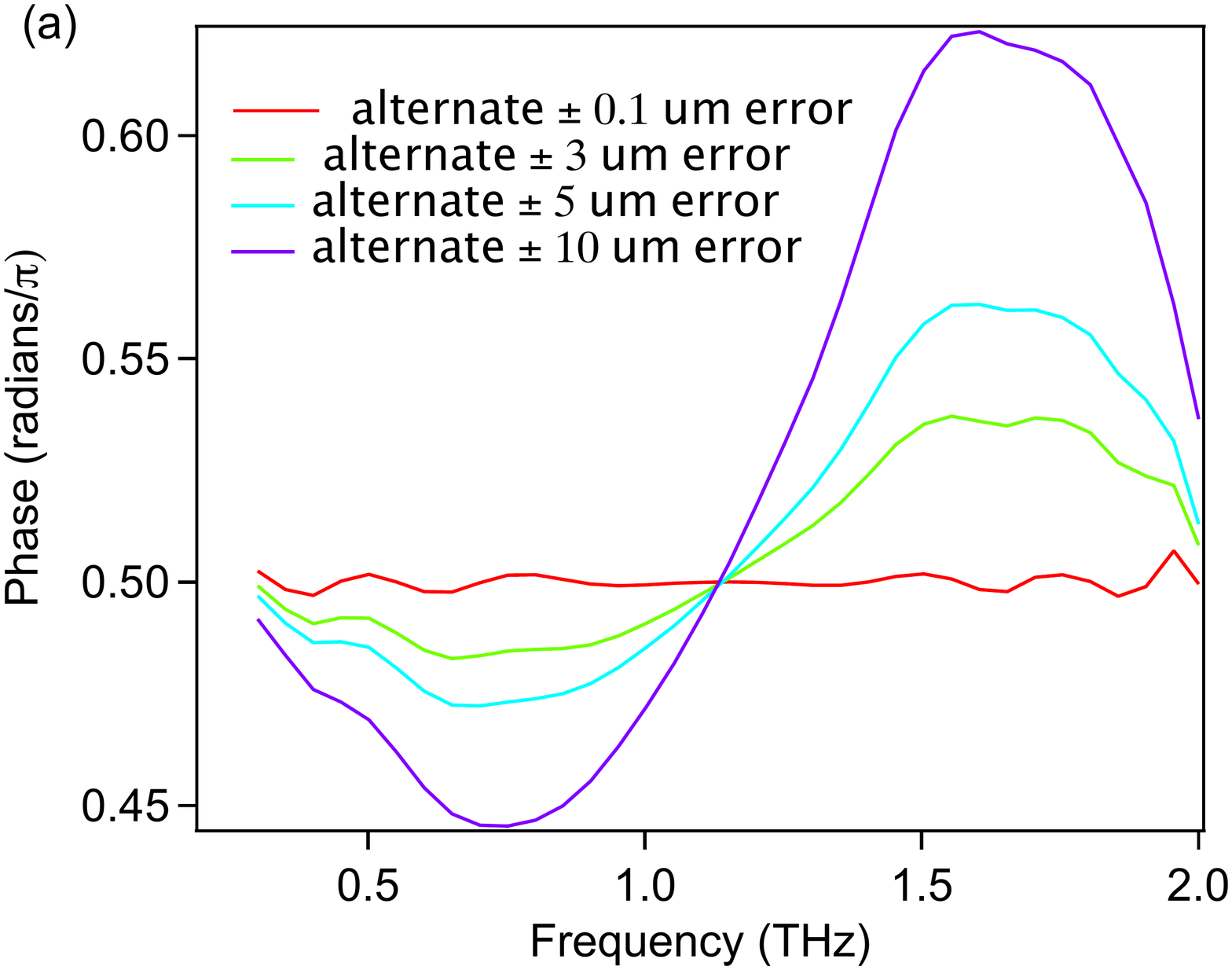}
\includegraphics[angle = -0,width=8.5cm]{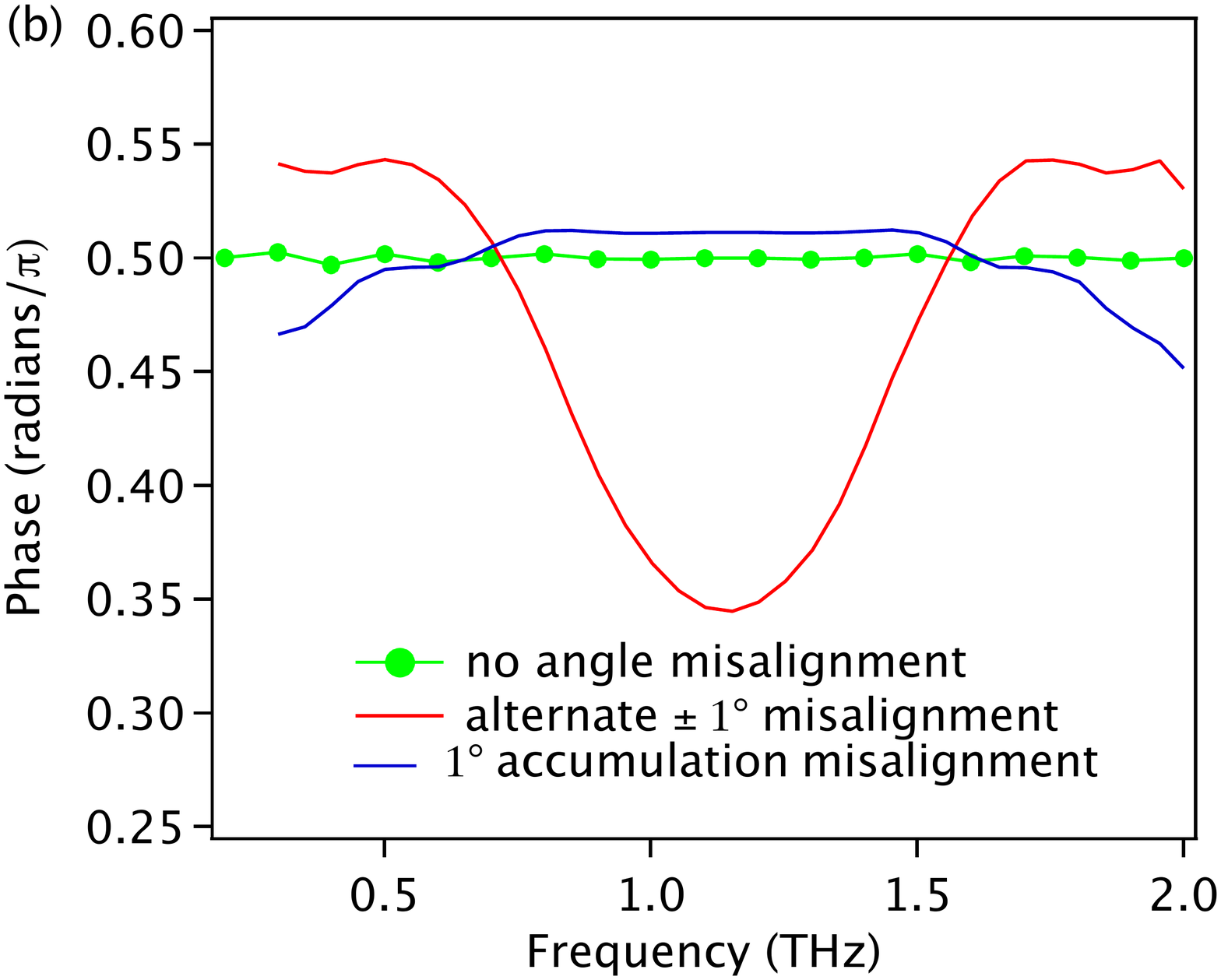}

\caption{Simulated phase of configuration C with simulated error for two different kinds of errors.  (a)  Red, blue, green and purple signify the phase delays when offsetting each sapphire layer by $\pm 0.1  \mu m$, $\pm 3  \mu m$,  $\pm 5  \mu m$ and $\pm 10  \mu m$ alternately from its ideal thickness. (b)  Red signifies offsetting each sapphire layer by $\pm 1  ^\circ$ alternately from its ideal rotation.   Blue signifies an accumulation of errors with each layer $+1 ^\circ$  additional degree per layer from its ideal position. }
\label{Simulation2}
\end{figure} 

Although the large birefringent contrast in sapphire allows us to shrink the overall length of the device by approximately a factor of 10, it also significantly decreases the tolerances for machining and assembly errors to achieve a particular phase error. In Fig. \ref{Simulation2} (a), we show the calculated phase error when thickness deviations of $\pm 0.1 \mu m$, $\pm 3 \mu m$, $\pm 5 \mu m$, and $\pm 10 \mu m$ from the ideal are substituted alternately in the stack.  In Fig. \ref{Simulation2} (b), we show the calculated phase error for $\pm 1 ^\circ$ misalignments for configuration C in the optimal thicknesses for sapphire as well as a model with a cumulative angular offset error where each layer is offset by an additional $ 1 ^\circ$ from the adjacent layer.   These simulations show that the device must be assembled very precisely to not have substantial phase errors.

Note that when comparing Fig. \ref{Simulation1} a and b, the two-thickness configuration for 8 plates has much smaller phase error than the single thickness for 12 plates. In fact, in order to reach 0.5 $\%$ error for one thickness, one needs at least 18 plates. We also simulated the case of different thickness for each plate up to 20 plates and we got similar 0.5 $\%$ error.  However, considering the extreme sensitivity of angular alignment, we believe it will be most suitable to pursue a design with fewer plates. Therefore, in the interest of simplicity and low manufacturing cost we chose to only have the single configuration C milled, as milling a number of plates of the same thickness is most cost effective.  The sapphire discs were machined with $A$ plane orientation to the specified target thicknesses with a diameter of 25.4 mm by Base Optics Inc.  Due to the large contrast in the indices of refraction between the extraordinary and ordinary direction, in principle we required that double-side polished sapphire plates must be machined to better than 1 $\mu$m tolerances.   And although a ``best effort" basis to achieve 1 $\mu$m was attempted, in practice only 10 $\mu$m could be guaranteed.  To compensate for this, a large number of plates were machined, all had their thicknesses measured with a micrometer with accuracy of 1 $\mu$m, and the ones closest to our specification were chosen.   We found than plates were usually within $\pm$3 $\mu$m of tolerance.  After choosing plates, we reoptimized the rotations through the basin-hopping algorithm with the thickness fixed at the measured values.   This achieved a predicted phase shift that was within 0.5$\%$ percent in a frequency range of 0.2 to 2 THz, otherwise the phase errors are expected to be large.

\begin{figure}[tbp]
\centering
\includegraphics[angle = 0 , width=4cm]{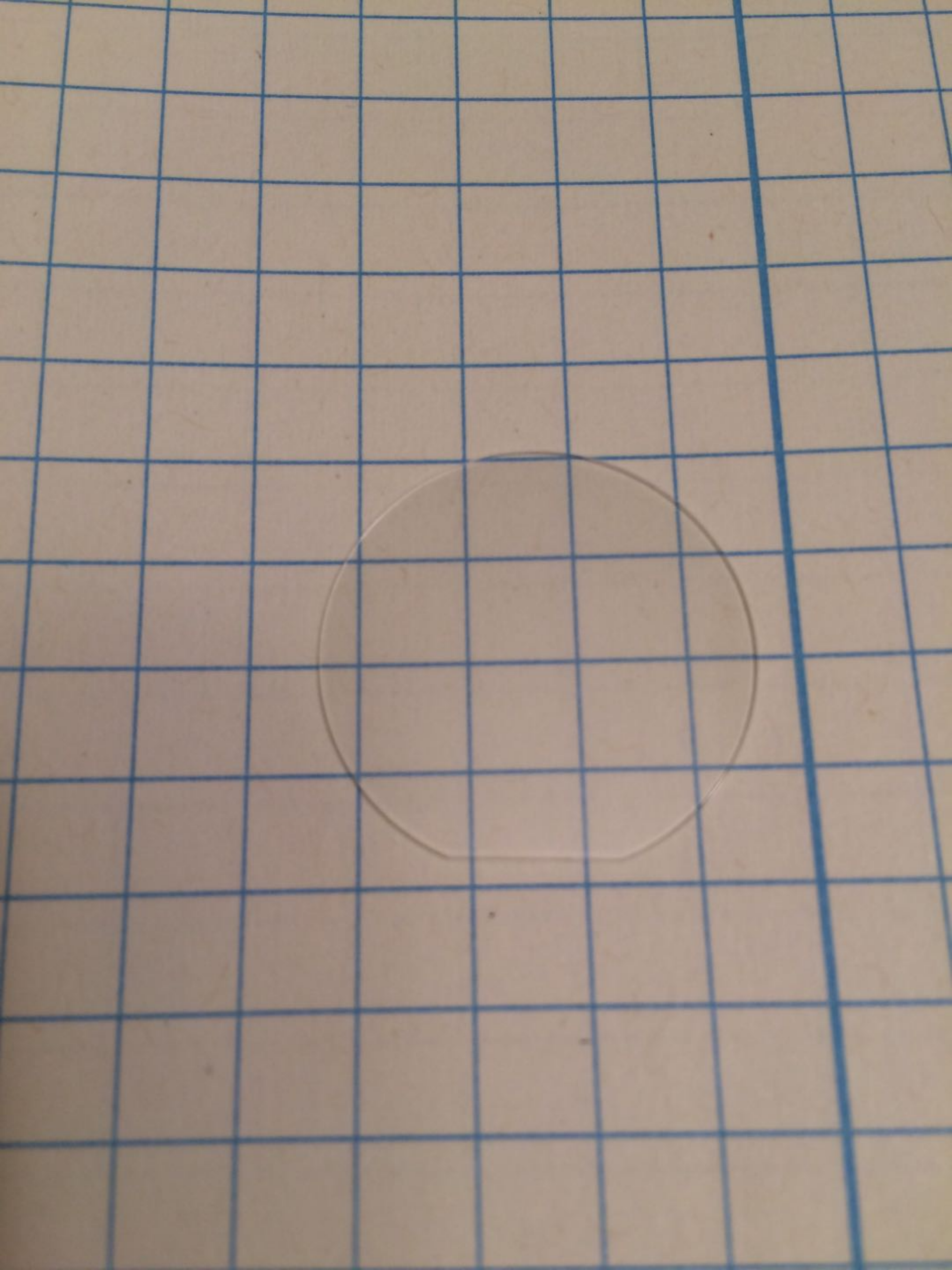}
\includegraphics[angle = 0 , width=3.65cm]{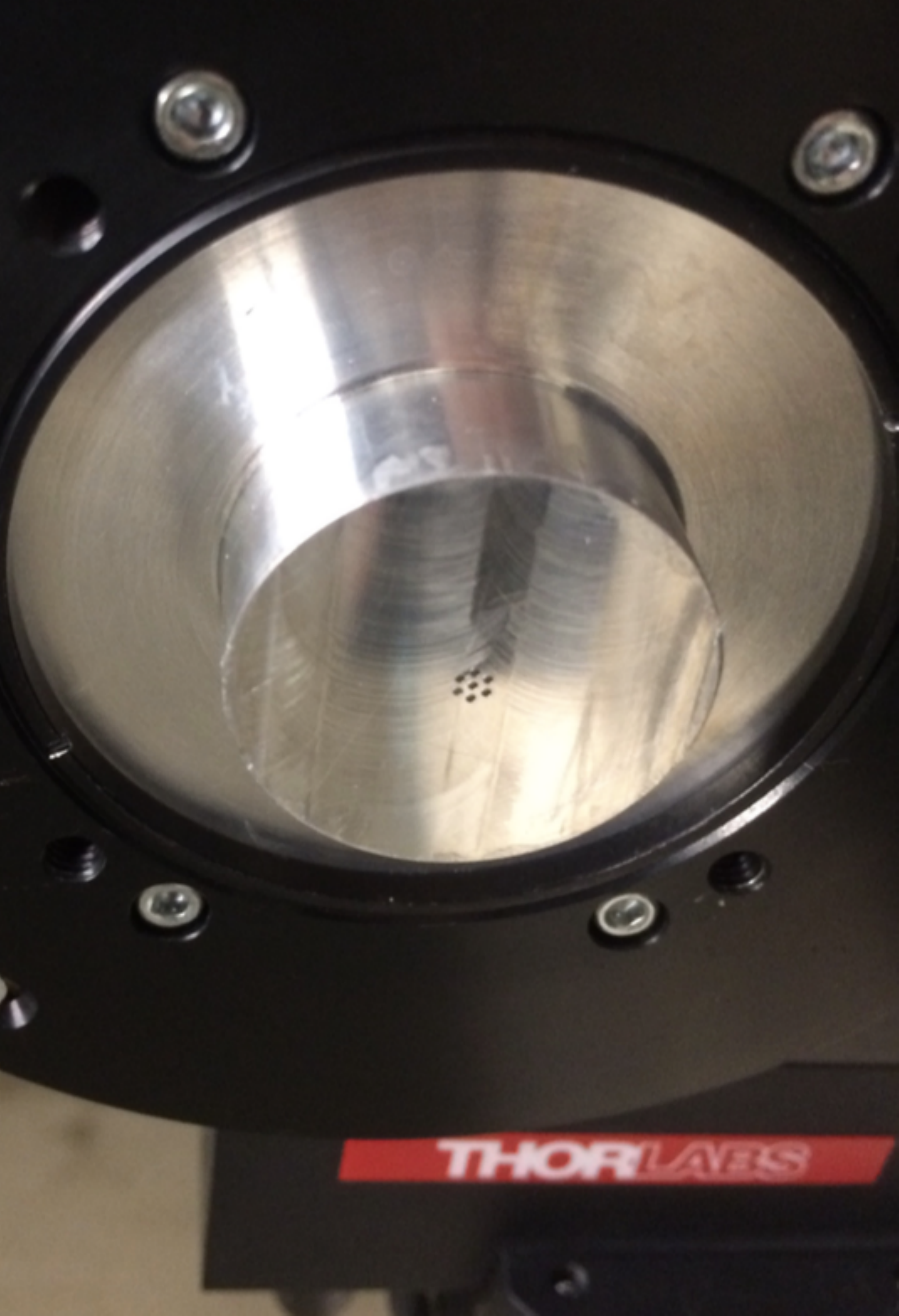}
\caption{a) View of a sapphire disc.   b) Face of jig that can apply vacuum to hold a sapphire disc on while it pressed in position and epoxied.  Note the small flat on the right side of the image that mates with a machined flat on the sapphire discs to reference the disc optical axis.}
\label{Pictures}
\end{figure}

A small flat was machined on each disc to reference the extraordinary axis.   Please see an example disc in Fig. \ref{Pictures}.  A small jig was devised that mated to this flat.  See Fig. \ref{Pictures} for a picture of the setup.  On the face of the jig were a few small holes in which suction could be applied while a disc was affixed and pressed into contact with other discs.   Each disc was epoxied on its outside edge with a small unpolished bevel on the discs aiding in epoxy adhesion.

THz measurements was performed in a home built THz setup using the polarization modulation technique developed recently \cite{aguilar2012terahertz,MorrisOE12,George12a}.  GaAs Auston switches are used as emitters and receivers to generate and detect THz pulses. An ultrafast laser (800 nm) is split into two paths by a beamsplitter. One beam travels to the biased emitter and generates a THz pulse. This THz pulse passes through the sample or substrate and arrives at the receiver. The other laser beam propagates to the receiver and is used to gate the THz pulse coming from the emitter side.   By varying the path length difference of the two paths, the electric field of the transmitted pulse is measured in the time domain.  Ratioing the Fourier transform of the transmission through the sample to that of a reference gives the frequency dependent complex transmission function in a range that typically of order 100 GHz - 3 THz. The complex optical parameters of interest e.g. complex dielectric function or the complex index of refraction can then be obtained from the complex transmission function. 

We use the fast polarization modulation technique to measure the polarization states accurately \cite{aguilar2012terahertz,MorrisOE12,George12a}.  A static wire-grid polarizer (WGP1) is placed before the sample. After the sample, a fast rotating polarizer (FRP) unit and then another static WGP2 are used. WGP1 and WGP2 transmit vertically polarized light. With this combination, in the phase modulation technique, $E_{x}(t)$ and $E_{y}(t)$ can be measured simultaneously in a single scan by reading off the in- and out-of-phase outputs from a lockin.   If light is originally sent in along the $x$ direction, then the complex rotation $\theta=\theta^{'}+i\theta^{''}$ can be obtained by $\theta$=atan[$E_{y}$($\omega$)/$E_{x}$($\omega$)] after Fourier transforming into the frequency domain. Linearly polarized light becomes elliptically polarized after passing through the sample with a rotation $\theta^{'}$ (real part). The imaginary part of the rotation $\theta^{''}$ is related to the ellipticity in the small rotation angle regime \cite{MorrisOE12}.

\begin{figure}[tbp]
\centering
\includegraphics[width=8.5cm]{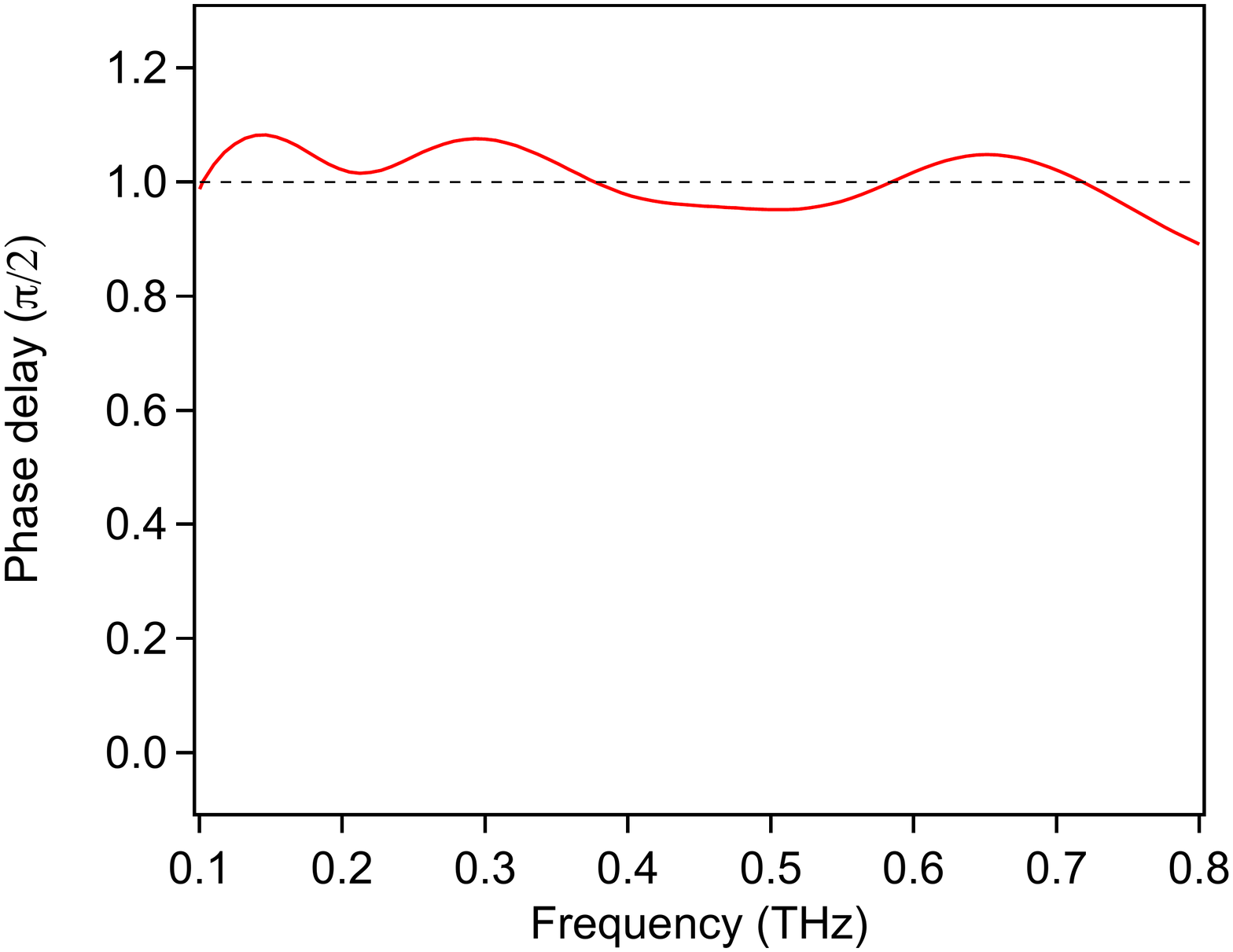}
\includegraphics[width=8.5cm]{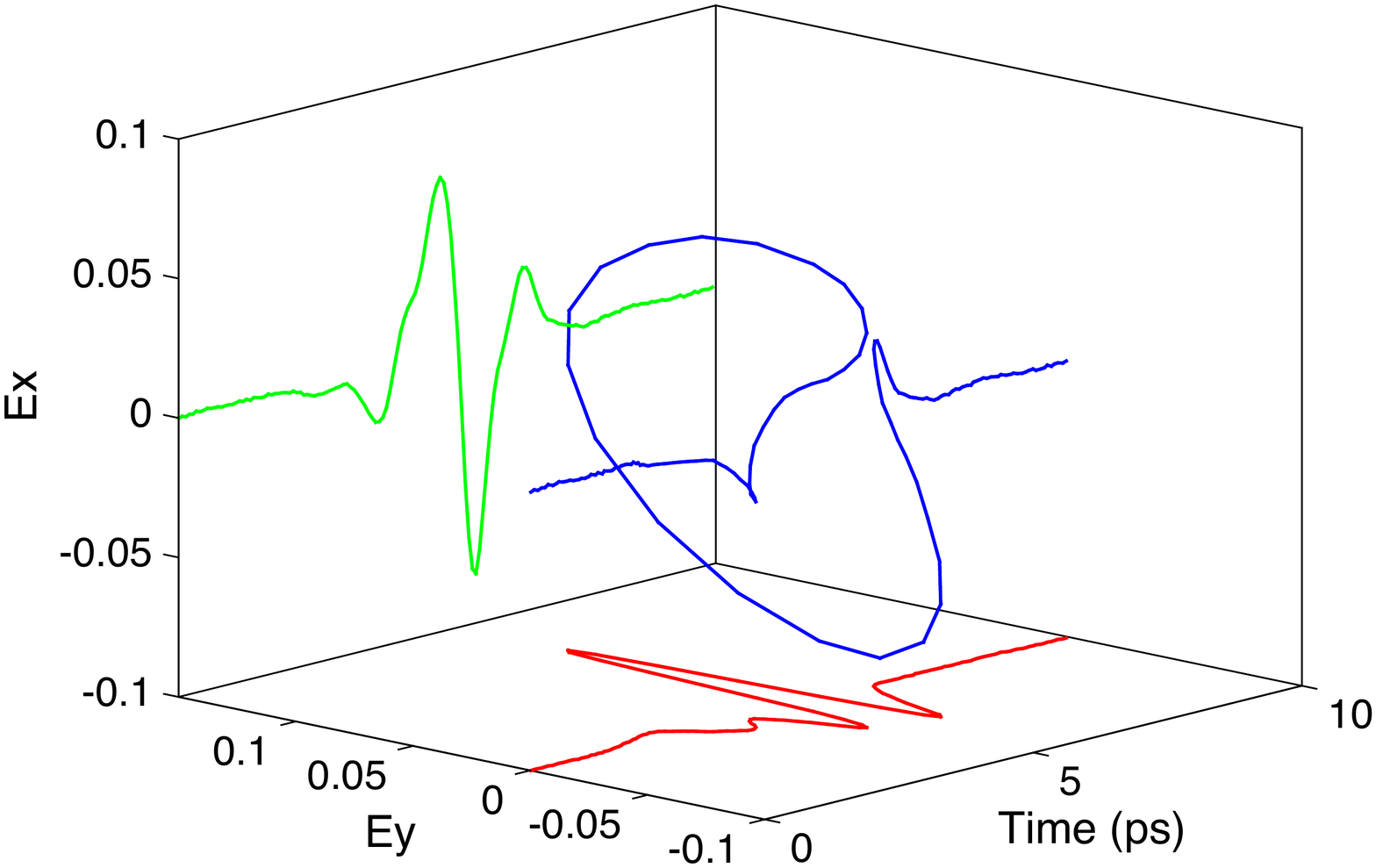}
\caption{ (a) Phase delay with incident electric field at 45$^{\circ}$ to the polarizer optical axis. (b) How the circularly polarized electric field evolves with the time.}
\label{Phaseresults}
\end{figure}

In Fig.  \ref{Phaseresults} we show the arctan$(E_y/E_x)$ which gives the relative $\sim \pi/2$  phase between the $x$ and $y$ channels and we also show how the circularly polarized electric field evolves with time.    Although the spectral range is more limited than in the prediction,  one can see errors of only a few percent (a standard deviation 4.7 $\%$) over the large frequency range of the 0.1  - 0.8 THz range.   Errors are larger  than the phase accuracy of spectrometer (standard deviation 1.3 $\%$).  These larger than anticipated phase errors could arise from errors in assembly alignment,  machining, or a small deviation of the used values from the actual values in the index of refraction of sapphire\cite{Neshat12a}.

\section{\label{Conclusion}Conclusion}

We have demonstrated the design and characterization of a THz range achromatic quarter-wave plate based on a stack of aligned variable thickness birefringent sapphire discs.   We determined the disc thicknesses and relative rotations through a basin-hopping Monte Carlo thermal annealing routine.  The basin-hopping scheme allows an improved refinement of the required thicknesses and rotations to give a phase error from the ideal $\pi/2$ of only $0.5 \%$, which is a factor of approximately 6 better than previous efforts. In fact, this value of below the standard spectrometer measurement accuracy of $\sim$ 1 $\%$. The large contrast between extraordinary and ordinary axes of sapphire allow us to greatly decrease the overall optical path length of our wave plate design by approximately a factor of 10 over existing similar designs based on quartz discs.  However, this very same contrast requires very precise tolerances in the milled thicknesses of the discs and their assembly.   We have detailed a method to compensate for differences in the thickness from their calculated ideal values.   We have constructed one of our designs and found it similar in performance to existing schemes, but using our much more compact geometry.

\section{\label{Acknowledgements}Acknowledgements}

We thank M. Neshat for sharing data on the index of refraction of sapphire. The THz instrumentation development was funded by the Gordon and Betty Moore Foundation through Grant GBMF2628 to NPA and the NSF through DMR-1508645. T.M. was supported by NSF DMR-1352373.

\bibliographystyle{unsrt}
\bibliography{mybib}{}

\begin{thebibliography}{10}

\bibitem{nuss1998terahertz}
Martin~C. Nuss and Joseph Orenstein.
\newblock Terahertz time-domain spectroscopy.
\newblock In {\em Millimeter and submillimeter wave spectroscopy of solids},
  pages 7--50. Springer, 1998.

\bibitem{kaindl2003ultrafast}
Robert~A. Kaindl, Marc~A. Carnahan, D.~H{\"a}gele, R.~L{\"o}venich, and
  Daniel~S. Chemla.
\newblock Ultrafast terahertz probes of transient conducting and insulating
  phases in an electron--hole gas.
\newblock {\em Nature}, 423(6941):734--738, 2003.

\bibitem{heyman1998time}
James~N. Heyman, Roland Kersting, and Karl Unterrainer.
\newblock Time-domain measurement of intersubband oscillations in a quantum
  well.
\newblock {\em Applied physics letters}, 72(6):644--646, 1998.

\bibitem{richter2010exciton}
Christiaan Richter and Charles~A. Schmuttenmaer.
\newblock {Exciton-like trap states limit electron mobility in TiO$_2$
  nanotubes}.
\newblock {\em Nature Nanotechnology}, 5(11):769--772, 2010.

\bibitem{taday2004applications}
Philip~F. Taday.
\newblock Applications of terahertz spectroscopy to pharmaceutical sciences.
\newblock {\em Philosophical Transactions of the Royal Society of London A:
  Mathematical, Physical and Engineering Sciences}, 362(1815):351--364, 2004.

\bibitem{choi2004potential}
Min~Ki Choi, Alan Bettermann, and D.W. Van Der~Weide.
\newblock Potential for detection of explosive and biological hazards with
  electronic terahertz systems.
\newblock {\em Philosophical Transactions of the Royal Society of London A:
  Mathematical, Physical and Engineering Sciences}, 362(1815):337--349, 2004.

\bibitem{leahy2007wideband}
M.R. Leahy-Hoppa, M.J. Fitch, X.~Zheng, L.M. Hayden, and R.~Osiander.
\newblock Wideband terahertz spectroscopy of explosives.
\newblock {\em Chemical Physics Letters}, 434(4):227--230, 2007.

\bibitem{hu1995imaging}
B.B. Hu and M.C. Nuss.
\newblock Imaging with terahertz waves.
\newblock {\em Optics letters}, 20(16):1716--1718, 1995.

\bibitem{johnson2001enhanced}
Jon~L. Johnson, Timothy~D. Dorney, and Daniel~M. Mittleman.
\newblock Enhanced depth resolution in terahertz imaging using phase-shift
  interferometry.
\newblock {\em Applied Physics Letters}, 78(6):835--837, 2001.

\bibitem{mittleman1997t}
Daniel~M. Mittleman, Stefan Hunsche, Luc Boivin, and Martin~C. Nuss.
\newblock T-ray tomography.
\newblock {\em Optics letters}, 22(12):904--906, 1997.

\bibitem{wang2003t}
S.~Wang, B.~Ferguson, D.~Abbott, and X.-C. Zhang.
\newblock T-ray imaging and tomography.
\newblock {\em Journal of Biological Physics}, 29(2-3):247--256, 2003.

\bibitem{chan2007imaging}
Wai~Lam Chan, Jason Deibel, and Daniel~M. Mittleman.
\newblock Imaging with terahertz radiation.
\newblock {\em Reports on progress in physics}, 70(8):1325, 2007.

\bibitem{corson1999vanishing}
John Corson, R.~Mallozzi, J.~Orenstein, J.N. Eckstein, and I.~Bozovic.
\newblock Vanishing of phase coherence in underdoped
  {Bi$_2$Sr$_2$CaCu$_2$O$_{8+\delta}$}.
\newblock {\em Nature}, 398(6724):221--223, 1999.

\bibitem{bilbro2011temporal}
L.~S. Bilbro, R.~Vald{\'e}s Aguilar, G.~Logvenov, O.~Pelleg, I.~Bozovi{\'c},
  and N.~P. Armitage.
\newblock Temporal correlations of superconductivity above the transition
  temperature in {La$_{2-x}$Sr$_x$CuO$_4$} probed by terahertz spectroscopy.
\newblock {\em Nature Physics}, 7(4):298--302, 2011.

\bibitem{aguilar2012terahertz}
R.~Vald{\'e}s Aguilar, A.~V. Stier, W.~Liu, L.S. Bilbro, D.~K. George,
  N.~Bansal, L.~Wu, J.~Cerne, A.G. Markelz, S.~Oh, et~al.
\newblock {Terahertz response and colossal Kerr rotation from the surface
  states of the topological insulator {Bi$_2$Se$_3$}}.
\newblock {\em Physical review letters}, 108(8):087403, 2012.

\bibitem{aguilar2013aging}
R.~Vald{\'e}s Aguilar, L.~Wu, A.V. Stier, L.S. Bilbro, M.~Brahlek, N.~Bansal,
  S.~Oh, and N.P. Armitage.
\newblock Aging and reduced bulk conductance in thin films of the topological
  insulator {Bi$_2$Se$_3$}.
\newblock {\em Journal of Applied Physics}, 113(15):153702, 2013.

\bibitem{wu2013sudden}
Liang Wu, M.~Brahlek, R.~Vald{\'e}s Aguilar, A.V. Stier, C.M. Morris,
  Y.~Lubashevsky, L.S. Bilbro, N.~Bansal, S.~Oh, and N.P. Armitage.
\newblock A sudden collapse in the transport lifetime across the topological
  phase transition in {(Bi$_{1-x}$In$_x$)$_2$Se$_3$}.
\newblock {\em Nature Physics}, 9(7):410--414, 2013.

\bibitem{hancock2011surface}
Jason~N. Hancock, Jacobus Lodevicus~Martinu van Mechelen, Alexey~B. Kuzmenko,
  Dirk Van Der~Marel, Christoph Br{\"u}ne, Elena~G. Novik, Georgy~V. Astakhov,
  Hartmut Buhmann, and Laurens~W. Molenkamp.
\newblock Surface state charge dynamics of a high-mobility three-dimensional
  topological insulator.
\newblock {\em Physical review letters}, 107(13):136803, 2011.

\bibitem{pan2014low}
LiDong Pan, Se~Kwon Kim, A~Ghosh, Christopher~M. Morris, Kate~A. Ross, Edwin
  Kermarrec, Bruce~D. Gaulin, S.M. Koohpayeh, Oleg Tchernyshyov, and N.P.
  Armitage.
\newblock Low-energy electrodynamics of novel spin excitations in the quantum
  spin ice {Yb$_2$Ti$_2$O$_7$}.
\newblock {\em Nature communications}, 5, 2014.

\bibitem{morris2014hierarchy}
C.M. Morris, R.~Vald{\'e}s Aguilar, A.~Ghosh, S.M. Koohpayeh, J.~Krizan, R.J.
  Cava, O.~Tchernyshyov, T.M. McQueen, and N.P. Armitage.
\newblock Hierarchy of bound states in the one-dimensional ferromagnetic ising
  chain {CoNb$_2$O$_6$} investigated by high-resolution time-domain terahertz
  spectroscopy.
\newblock {\em Physical review letters}, 112(13):137403, 2014.

\bibitem{laurita2015singlet}
N.J. Laurita, J.~Deisenhofer, LiDong Pan, C.M. Morris, M.~Schmidt, M.~Johnsson,
  V.~Tsurkan, A.~Loidl, and N.P. Armitage.
\newblock {Singlet-Triplet Excitations and Long-Range Entanglement in the
  Spin-Orbital Liquid Candidate FeSc$_2$S$_4$}.
\newblock {\em Physical Review Letters}, 114(20):207201, 2015.

\bibitem{bosse2014a}
G.~Boss{\'e}, L.S. Bilbro, R.~Vald{\'e}s Aguilar, LiDong Pan, Wei Liu, A.V.
  Stier, Y.~Li, L.H. Greene, J.~Eckstein, and N.P. Armitage.
\newblock Low energy electrodynamics of the kondo-lattice antiferromagnet
  cecu$_2$ge$_2$.
\newblock {\em Physical Review B}, 85(15):155105, 2012.

\bibitem{little2017antiferromagnetic}
A.~Little, Liang Wu, P.~Lampen-Kelley, A.~Banerjee, S.~Patankar, D.~Rees, C.A.
  Bridges, J.-Q. Yan, D.~Mandrus, S.E. Nagler, et~al.
\newblock {Antiferromagnetic Resonance and Terahertz Continuum in
  $\alpha$-RuCl$_3$}.
\newblock {\em Physical Review Letters}, 119(22):227201, 2017.

\bibitem{Xia06a}
Jing Xia, Yoshiteru Maeno, Peter~T. Beyersdorf, M.M. Fejer, and Aharon
  Kapitulnik.
\newblock High resolution polar kerr effect measurements of sr 2 ruo 4:
  evidence for broken time-reversal symmetry in the superconducting state.
\newblock {\em Physical review letters}, 97(16):167002, 2006.

\bibitem{Qi2008}
Xiao-Liang Qi, Taylor~L. Hughes, and Shou-Cheng Zhang.
\newblock Topological field theory of time-reversal invariant insulators.
\newblock {\em Physical Review B}, 78(19):195424, 2008.

\bibitem{Nandkishore2012}
Rahul Nandkishore, L.S. Levitov, and A.V. Chubukov.
\newblock Chiral superconductivity from repulsive interactions in doped
  graphene.
\newblock {\em Nature Physics}, 8(2):158--163, 2012.

\bibitem{Tse11a}
Wang-Kong Tse and A.H. MacDonald.
\newblock Magneto-optical faraday and kerr effects in topological insulator
  films and in other layered quantized hall systems.
\newblock {\em Physical Review B}, 84(20):205327, 2011.

\bibitem{Maciejko10a}
Joseph Maciejko, Xiao-Liang Qi, H.~Dennis Drew, and Shou-Cheng Zhang.
\newblock Topological quantization in units of the fine structure constant.
\newblock {\em Physical review letters}, 105(16):166803, 2010.

\bibitem{Armitage14a}
N.P. Armitage.
\newblock {Constraints on Jones transmission matrices from time-reversal
  invariance and discrete spatial symmetries}.
\newblock {\em Physical Review B}, 90(3):035135, 2014.

\bibitem{Castro-Camus2005}
Enrique Castro-Camus.
\newblock Polarization-resolved terahertz time-domain spectroscopy.
\newblock {\em Journal of Infrared, Millimeter, and Terahertz Waves},
  33(4):418--430, 2012.

\bibitem{Makabe2007}
Hiroyuki Makabe, Yuichi Hirota, Masahiko Tani, and Masanori Hangyo.
\newblock Polarization state measurement of terahertz electromagnetic radiation
  by three-contact photoconductive antenna.
\newblock {\em Optics express}, 15(18):11650--11657, 2007.

\bibitem{Shimano2011}
R.~Shimano, Y.~Ikebe, K.S. Takahashi, M.~Kawasaki, N.~Nagaosa, and Y.~Tokura.
\newblock Terahertz faraday rotation induced by an anomalous hall effect in the
  itinerant ferromagnet srruo$_3$.
\newblock {\em EPL (Europhysics Letters)}, 95(1):17002, 2011.

\bibitem{Neshat12a}
M.~Neshat and N.P. Armitage.
\newblock Improved measurement of polarization state in terahertz polarization
  spectroscopy.
\newblock {\em Optics letters}, 37(11):1811--1813, 2012.

\bibitem{MorrisOE12}
Christopher~M. Morris, R.~Vald{\'e}s Aguilar, Andreas~V. Stier, and N~Peter
  Armitage.
\newblock Polarization modulation time-domain terahertz polarimetry.
\newblock {\em Optics express}, 20(11):12303--12317, 2012.

\bibitem{George12a}
Deepu~K. George, Andreas~V. Stier, Chase~T. Ellis, Bruce~D. McCombe, John
  {\v{C}}erne, and Andrea~G. Markelz.
\newblock Terahertz magneto-optical polarization modulation spectroscopy.
\newblock {\em JOSA B}, 29(6):1406--1412, 2012.

\bibitem{koirala2015record}
Nikesh Koirala, Matthew Brahlek, Maryam Salehi, Liang Wu, Jixia Dai, Justin
  Waugh, Thomas Nummy, Myung-Geun Han, Jisoo Moon, Yimei Zhu, et~al.
\newblock Record surface state mobility and quantum hall effect in topological
  insulator thin films via interface engineering.
\newblock {\em Nano letters}, 15(12):8245--8249, 2015.

\bibitem{wu2015high}
Liang Wu, Wang-Kong Tse, M.~Brahlek, C.M. Morris, R.~Vald{\'e}s Aguilar,
  N.~Koirala, S.~Oh, and N.P. Armitage.
\newblock {High-resolution Faraday rotation and electron-phonon coupling in
  surface states of the bulk-insulating topological insulator
  Cu$_{0.02}$Bi$_2$Se$_3$}.
\newblock {\em Physical review letters}, 115(21):217602, 2015.

\bibitem{wu2016tuning}
Liang Wu, R.M. Ireland, M.~Salehi, B.~Cheng, N.~Koirala, S.~Oh, H.E. Katz, and
  N.P. Armitage.
\newblock {Tuning and stabilizing topological insulator Bi$_2$Se$_3$ in the
  intrinsic regime by charge extraction with organic overlayers}.
\newblock {\em Applied Physics Letters}, 108(22):221603, 2016.

\bibitem{wu2016quantized}
Liang Wu, M.~Salehi, N.~Koirala, J.~Moon, S.~Oh, and N.P. Armitage.
\newblock {Quantized Faraday and Kerr rotation and axion electrodynamics of a
  3D topological insulator}.
\newblock {\em Science}, 354(6316):1124--1127, 2016.

\bibitem{laurita2017anomalous}
N.~J. Laurita, Yi~Luo, Rongwei Hu, Meixia Wu, S.~W. Cheong, O.~Tchernyshyov,
  and N.~P. Armitage.
\newblock {Asymmetric Splitting of an Antiferromagnetic Resonance via Quartic
  Exchange Interactions in Multiferroic Hexagonal ${\mathrm{HoMnO}}_{3}$}.
\newblock {\em Phys. Rev. Lett.}, 119:227601, Dec 2017.

\bibitem{West49a}
C.D. West and A.S. Makas.
\newblock The spectral dispersion of birefringence, especially of birefringent
  plastic sheets.
\newblock {\em JOSA}, 39(9):791--794, 1949.

\bibitem{Destriau49a}
G.~Destriau and J.~Prouteau.
\newblock {R{\'e}alisation d'un quart d'onde quasi achromatique par
  juxtaposition de deux lames cristallines de m{\^e}me nature}.
\newblock {\em J. Phys. Radium}, 10(2):53--55, 1949.

\bibitem{Masson06a}
Jean-Baptiste Masson and Guilhem Gallot.
\newblock Terahertz achromatic quarter-wave plate.
\newblock {\em Optics letters}, 31(2):265--267, 2006.

\bibitem{wales1997global}
David~J. Wales and Jonathan~P.K. Doye.
\newblock {Global optimization by basin-hopping and the lowest energy
  structures of Lennard-Jones clusters containing up to 110 atoms}.
\newblock {\em The Journal of Physical Chemistry A}, 101(28):5111--5116, 1997.

\end{thebibliography}

\end{document}